\documentclass[12pt,a4paper]{article}
\usepackage{lscape}
\usepackage{xcolor}
\usepackage{graphicx}
\usepackage[utf8]{inputenc}
\usepackage[T1]{fontenc}
\usepackage[english]{babel}
\usepackage{eso-pic}
\usepackage{mathrsfs}
\usepackage{url}
\usepackage{amssymb}
\usepackage{amsmath}
\usepackage{multirow}
\usepackage{hyperref}
\usepackage{booktabs}
\usepackage{rotating,capt-of}
\usepackage{makeidx}
\usepackage{subfigure}
\usepackage{bbm,cases}
\usepackage{cooltooltips,epstopdf}
\usepackage{tablefootnote}
\usepackage{float}
\usepackage{eurosym}
\usepackage[style=authoryear, backend=biber, maxcitenames=2, maxbibnames=99]{biblatex}
\newcommand{\mycite}[1]{\textcolor{blue}{\cite{#1}}}
\usepackage{textcomp}
\usepackage{comment}
\usepackage{siunitx}
\makeatletter
\newcommand{\Kommentar}[1]{}

\usepackage{ntheorem}
\usepackage{xcolor}
\definecolor{beige}{rgb}{0.96,0.96,0.863}
\theoremstyle{nonumberplain}
\usepackage[framemethod=tikz]{mdframed}
\newmdtheoremenv[backgroundcolor=beige,%
outerlinewidth=0.5pt,roundcorner=8pt,leftmargin=0pt,rightmargin=0pt,%
outerlinecolor=black,innertopmargin=5pt,splittopskip=\topskip,%
ntheorem=true]{result}{Result}

\begin{document}

\title{Bridging an energy system model with an ensemble deep-learning approach for electricity price forecasting}

\author{Souhir Ben Amor\footnote{Brandenburgische Technische Universit\"at Cottbus-Senftenberg
Chair of Energy Economics, Email: benamor@b-tu.de}
\and Thomas M\"obius\footnote {Brandenburgische Technische Universit\"at Cottbus-Senftenberg
Chair of Energy Economics, Email: Thomas.Moebius@b-tu.de} \and Felix M\"usgens\footnote{Brandenburgische Technische Universit\"at Cottbus-Senftenberg
Chair of Energy Economics, Email: felix.muesgens@b-tu.de} 
}

\date{\today}
\maketitle

\begin{abstract}
\noindent 
This paper combines a techno-economic energy system model with an econometric model to maximise electricity price forecasting accuracy. The proposed combination model is tested on the German day-ahead wholesale electricity market. Our paper also benchmarks the results against several econometric alternatives. Lastly, we demonstrate the economic value of improved price estimators maximising the revenue from an electric storage resource.\\
The results demonstrate that our integrated model improves overall forecasting accuracy by 18\%, compared to available literature benchmarks. Furthermore, our robustness checks reveal that a) the Ensemble Deep Neural Network model performs best in our dataset and b) adding output from the techno-economic energy systems model as econometric model input improves the performance of all econometric models.\\
The empirical relevance of the forecast improvement is confirmed by the results of the exemplary storage optimisation, in which the integration of the techno-economic energy system model leads to a revenue increase of up to 10\%.

\end{abstract}
\normalsize{\textbf{Keywords}: Electricity price forecasting, Econometric price forecasts, Techno-economic models, Ensemble Deep Neural Network model, Energy trading, Storage bidding strategy.\\

}
\textbf{JEL classification:} C52, C53, Q40, D44.

\section{Introduction and Motivation}

Producers and consumers in many electricity systems worldwide are trading on liquid wholesale markets. Prices are vital in coordinating supply and demand. Market participants need effective price forecasts to optimise their business strategies.\\ 
The forecasting of electricity prices focuses on either the short-term, with forecasting horizons in the order of hours or days, or the medium to long-term, in months or years (\mycite{weron2014electricity}). Fundamentally different types of models are used, mostly oriented around these time frames. Short-term price forecasting is predominantly approached with econometric methods, while long-term price forecasting is predominantly achieved using techno-economic models formulated as operations research-type optimisation problems. These techno-economic bottom-up models are often referred to as 'energy systems models' (ESM). Prices in ESM are estimated from dual variables (typically ''shadow prices'' of demand constraints, see e.g. \mycite{musgens2006quantifying}).\\
In this paper, we focus on short-term electricity price forecasting (EPF). Short-term EPF is challenging due to the complexity and the particular economic and physical characteristics of the electricity market, which account for the simultaneous occurrence of spikes, volatility, long memory, mean reversion, negative prices, and regime switches. Numerous researchers have devoted their time and expertise to these complexities and developed various econometric approaches (see \mycite{nowotarski2018recent} and \mycite{weron2014electricity} for an overview). \\
Although state-of-the-art econometric models incorporate explanatory variables such as electricity demand, fuel prices, and generation from renewable energy sources, they analyse market prices statistically, i.e. without explicitly modelling the underlying techno-economic objectives and constraints on the demand and supply sides, which in combination determine market prices.\\
In contrast, ESMs replicate price formation in electricity markets deciding on an efficient market equilibrium, taking into account the physical and dynamic properties of the market on both the supply and demand sides. ESMs play to their strengths in long-term forecasts but tend to be less effective when dealing with the formation of high-frequency price patterns, which in turn is a strength of econometric models. Nowadays, a small but growing number of researchers are interested in hybrid methods that combine ESM and econometric models for short-term price forecasting (e.g., \mycite{watermeyer2023hybrid}, \mycite{DEMARCOS2019240} and \mycite{MACIEJOWSKA20161051}). \\
Despite these first attempts, there exists a research gap in understanding whether results from a techno-economic energy system model add value to short-term price forecasts from a state-of-the-art econometric model. Our first objective in this paper is to close that gap. We combine an ESM and an econometric model to benefit from the respective strengths of both model types. For the proposed combination, both individual models should be among the best in their respective classes to achieve meaningful practical results. In this context, we select the \textit{em.power dispatch model}, a techno-economic dispatch ESM specifically tailored to provide day-ahead price forecasts in high temporal resolution. For the econometric model, we select the state-of-the-art Ensemble Deep Neural Network (Ens-DNN) model.\footnote {The choice of the Ens-DNN model is motivated by the results of \mycite{lago2021forecasting}, which demonstrate that the Ens-DNN model provides accurate forecasts. They also suggest it to benchmark results.} The link between the two models is the market clearing price (MCP). In the ESM, the MCP is the model output calculated as the shadow price of the demand constraint. In the econometric models, the MCP is an independent variable. Adding the MCP as an independent variable in the Ens-DNN improves forecasting accuracy by nearly 20\%, providing very accurate forecasts.\\
Our second objective is to assess the value of ESM input in econometric models in general. We provide quantitative results with an analysis of several other econometric models (ranging from simple to sophisticated). In addition to enabling a more general statement on the use of ESM results as input in econometric model forecasts, this also tests the relative forecast quality of the Ens-DNN model. We find that adding the ESM's MCP as an independent variable in econometric models improves forecast quality for all econometric models already performing reasonably well on their own.  Hence, this part of our work serves as a robustness check for the inclusion of ESM results as an independent variable in econometric models.\\ 
Our third contribution is to quantify the impact of the number of independent variables on forecasting accuracy and test the possibility of relying only on the ESM's MCP as a single regressor in econometric models. This also answers the question of to what extent the MCP computed by an ESM accommodates all the techno-economic variables usually taken into account in econometric models to capture electricity price variation.\\
Finally, we present a storage bidding strategy for the day-ahead market based on the forecasting results from the proposed model and the benchmark models that allow us to measure, in monetary terms, the economic effects of error prediction.\\
More accurate electricity price forecasts are relevant for all parties taking open positions in power markets: access to better day-ahead price forecasts than the competition directly increases market revenues. Furthermore, power generation companies and utilities benefit when making dispatch decisions, for single units as well as for portfolios of power plants and sector coupling technologies. On the demand side, our improved forecasts can be used to schedule flexible consumption at low-price hours. These examples suggest that forecasts constitute techno-economic inputs to energy company decision-making. The high demand for improved EPF models is therefore unquestionable.\\
The remainder of this paper is structured as follows: Section \ref{literature} summarises the related research on EPF and discusses the adopted methodology and the gaps left to be filled by our investigation. Section \ref{method} outlines the theoretical background of the ESM, the econometric models, and the combined model and explains the storage optimisation strategy. Section \ref{results} presents the case studies in which the proposed forecasting method has been tested. Complementary to the overall results, in section \ref{storage}, we discuss the implications of the forecast results for the storage bidding strategy. Subsequently, section \ref{conclusion} contains the conclusions drawn in this paper, incorporating recommendations for market participants.

\section{Related Literature}
\label{literature}

Forecasting in the electricity market has attracted academics, policymakers, and practitioners since the 1990s. Researchers have focused on developing reliable EPF models, and while statistical and econometric approaches are generally the dominant approach, long-term contexts also require a more fundamental understanding of market behaviour and dynamics (\mycite{weron2014electricity}).\\
Statistical models (\mycite{ben2018forecasting}, \mycite{amor2018forecasting}, \mycite{weron2008forecasting}, \mycite{ziel2016forecasting}, \mycite{lago2018forecasting}, and \mycite{conejo2005forecasting}) as well as artificial intelligence (AI) and machine learning (ML) methods  (\mycite{abedinia2015electricity},  \mycite{portoles2018electricity}, \mycite{lago2021forecasting}, and \mycite{jiang2018day}) have been widely accepted due to their ability to capture linear and non-linear trends, respectively. There have been studies that combine both approaches into a pure econometric hybrid model (\mycite{nikodinoska2022solar}, \mycite{chang2019electricity}) and \mycite{watermeyer2023hybrid}. \\
In addition, several researchers emphasise the importance of incorporating fundamental variables in econometric models when forecasting electricity prices (\mycite{karakatsani2008forecasting}, \mycite{kristiansen2012forecasting}, and \mycite{gianfreda2020comparing}).\\
Despite econometric models providing information about physical price drivers, for instance, demand, CO$_2$ prices, fuel prices, renewable energy, etc. (\mycite{karakatsani2008intra}), these models analyse market prices statistically and do not address the underlying demand and supply functions that drive market prices. Additionally, these models are usually built on historical data, which means that they can perform under the assumption that history will repeat itself in the future. However, in today's complex electricity markets, this is not always the case.
Hence, it is important to realise that although statistical models may be able to incorporate operations and market dynamics into their EPF, they often struggle to represent regulatory and market structural changes. \

To address these aspects, the ESM takes into consideration the various technical features of generation units and their production costs. More precisely, it simulates price formation from the intersection of demand with the physical and dynamic properties of the generators' supply functions, reflecting a more techno-economic "bottom-up" approach. As a result, the events that are difficult to model using statistics or econometrics are reflected in the estimated MCP of the ESM. \
Recently, the ESM has received much attention when it comes to deriving electricity price estimators. For example, several studies have been conducted in which competitive market prices have been modelled to investigate market power events (\mycite{musgens2006quantifying}, \mycite{Borenstein2002}). 
\mycite{SENSFU20083086} analyse the impact of renewable electricity generation on spot market prices, the so-called merit-order effect. \mycite{hirth2013market} apply an ESM to derive the market value of variable renewable energy.
Additionally, an advantage of ESMs is that they have the option to incorporate market participants' preferences, such as risk preferences (see e.g., \mycite{MOBIUS2023106767}).  \
However, despite being used widely in research, industry, and politics, ESMs are not the ideal tool for short-term price forecasting. This is because their performance when it comes to capturing short-term price dynamics is poor (\mycite {bello2016probabilistic}), while time-series econometric models are more accurate when it comes to providing short-term forecasts (\mycite{bunn2000forecasting}).\
Hence, there is a growing interest in combining ESMs with econometric methodologies to address their shortcomings and improve predictive performance.\
More precisely, this combination allows us to include fundamental market mechanisms and the most relevant economic drivers of electricity prices, such as demand, supply, and technical constraints, into the econometric model. This is in addition to the behavioural aspects (speculative and strategic behaviour) and statistical price features (spikes (\mycite{christensen2012forecasting}), high volatility, seasonality, etc.) provided by the econometric approach. Therefore, the behaviour and operation of the power market can be successfully incorporated into electricity price forecasts, which is of significant interest to its participants.\\
Such a combination has produced favourable outcomes for both point and probabilistic forecasting, particularly in medium-term scenarios (\mycite{watermeyer2023hybrid}, \mycite{bello2017medium}, \mycite{bello2016parametric}, \mycite{gonzalez2011forecasting}, and \mycite{karakatsani2008forecasting}). \\
Despite the importance of short-term price forecasting, short-term applications are very scarce and limited to the following studies: \mycite{de2019electricity} and \mycite{de2019short} propose a short-term hybrid EPF model, which combines a cost-production optimisation model with an Artificial Neural Network model to forecast the Iberian electricity market. In the same vein,  \mycite{gonzalez2011forecasting} combines a techno-economic model, formulated with supply stack modelling, with an econometric model (regime-switching model) using data on price drivers to predict the APX power exchange for Great Britain.\\
Considering the rarity of the published studies and the simplicity of the comparisons (limited to outdated methods and avoiding state-of-the-art econometric methods), the results cannot be generalised. In the preceding paragraphs, several deficiencies and scarcities have been highlighted in the context of short-term EPF, which motivates our proposed methodology.\

Price forecasting models that combine techno-economic optimisation models and econometric modelling approaches have shown promise over the medium term. Their performance in the short term, however, is rarely tested (see \mycite{watermeyer2023hybrid}). Therefore, it would be interesting to determine whether the same advantages can be achieved over the short term.\\
Moreover, another aspect of great importance is the way in which the techno-economic and econometric models are combined. The most widely used procedure is to obtain the MCP from the ESM, which is later included in the econometric model's input dataset. In previous research, the selection of the model components is neither analysed nor justified. In this regard, it is unclear to what extent the individual components are relevant or useful. As a result, it is imperative to carefully choose each model component. \\
Aside from this, the models used are outdated, including artificial neural networks and autoregressive models with exogenous variables (ARX) that have shown weaknesses in forecasting electricity prices (\mycite{lago2018forecasting}, \mycite{weron2014electricity}). \\
Firstly, one of the neural network's drawbacks is associated with the high computational training cost (\mycite{lago2018forecasting}). With the introduction of deep learning, this limitation has been overcome (\mycite{goodfellow2016deep}). Indeed, using the greedy layer-wise pretraining algorithm  \mycite{hinton2006fast} has demonstrated that deep belief networks can be efficiently trained, allowing researchers to train complex neural networks with more than one hidden layer. \\
To fill this gap in the literature, we propose to combine the ESM with the state-of-the-art Ensemble Deep Neural Network model, which is positioned at the forefront of electricity price forecasting models (\mycite{lago2021forecasting}).\\
Secondly, ARX is a linear model that incorporates an extensive number of input features. The traditional methods of estimating the ARX model (ordinary least square) are not able to deal with redundancy in the input features, resulting in unsatisfactory forecasting accuracy (\mycite{ziel2018day}, \mycite{uniejewski2018efficient}, and \mycite{ziel2016forecasting}). To address this restriction, use of the least absolute shrinkage and selection operator (LASSO) (\mycite{tibshirani1996regression}) and elastic nets (\mycite{zou2005regularization}) is proposed. These two regularisation strategies set some of the parameters to zero by jointly minimising the squared errors and a penalty factor of the model parameters. This results in the removal of redundant regressors. \\
In light of the benefits associated with LASSO techniques (\mycite{uniejewski2023lasso}, \mycite{uniejewski2016automated}), we have opted to employ the Ensemble LASSO Estimated AutoRegressive (Ens-LEAR) technique for benchmarking purposes.\\
This leads us to another issue when discussing combined approaches, in general, which is their comparison with benchmarks. In most research, advanced models are avoided (\mycite{Singh2018pso}, and \mycite{naz2019short}) or only outdated models are adopted for comparisons (\mycite{bento2018bat}, and \mycite{peesapati2017electricity}). In addition, a comparison of models that combine techno-economic and econometric approaches for short-term power price forecasting is restricted to their individual components and a na\"ive model. As a result, it is impossible to determine with precision how accurate the newly suggested approaches are. To fill this literature gap, a variety of econometric models commonly used in EPF, besides the Ens-LEAR model, were also adopted for comparison purposes. Each econometric model is combined separately with the techno-economic ESM, and their forecasting accuracy is then evaluated. This enables us to conduct a thorough evaluation, facilitating a robust analysis that effectively compares the performance of various approaches. This, in turn, enables us to confidently determine the robustness and generalisability of our methodology.

\section{Methodology}
\label{method}

In this section, we discuss our proposed forecasting methodology, which combines the strengths of both the techno-economic energy system model and the econometric model to improve electricity price forecasting. The methodology section starts with a comprehensive overview of the ESM. The second subsection presents the econometric models, presenting the Ensemble Deep Neural Network (Ens-DNN) model first (subsection \ref{Met_DNN}). We chose the Ens-DNN model as our first-choice econometric model because it is "highly accurate", based on findings from \mycite{lago2021forecasting},  who also suggest it as an open-source benchmark model for all "new complex EPF forecasting methods"  (\mycite{lago2021forecasting}). \\
We also combine the ESM with six other econometric models, which henceforth are referred to as 'benchmark models'. These benchmark models are described in subsection \ref{Met_econ}. Examining various econometric models in addition to the Ens-DNN holds significance for two key reasons: first, it benchmarks the forecast quality of the Ens-DNN model. Second, if including the MCP as an independent variable increases accuracy in different econometric models, our results can be considered more robust. The third subsection defines how the ESM and the econometric models are coupled.
In the last subsection, we present the methodology for the empirical valuation of the additional price forecast accuracy in an electricity storage optimisation problem.

\subsection{Energy System Model}
\label{fundm}

We apply the \textit{em.power dispatch model}\footnote{Model code and input data are available on GitHub: https://github.com/ProKoMoProject/Enhancing-Energy-System-Models-Using-Better-Load-Forecasts} as presented in \mycite{moebius2023enhancing}, which is formulated as a linear optimisation problem minimising overall system costs.
The model includes a comprehensive representation of key techno-economic elements of a liberalised European electricity sector, including international trade between markets, minimum generation and start-up restrictions on power plants, electricity generation from variable renewable energy sources, combined heat and power plants, energy storage, and control power provision. A demand constraint guarantees that supply and demand are balanced across all market zones and time periods. Following economic theory, the \textit{em.power dispatch model} calculates a market equilibrium in the electricity market, minimising total system costs. Wholesale electricity prices are computed during the optimisation as shadow prices of the demand constraint.

\begin{figure}[h]
  \centering
  \includegraphics[width=0.6\textwidth]{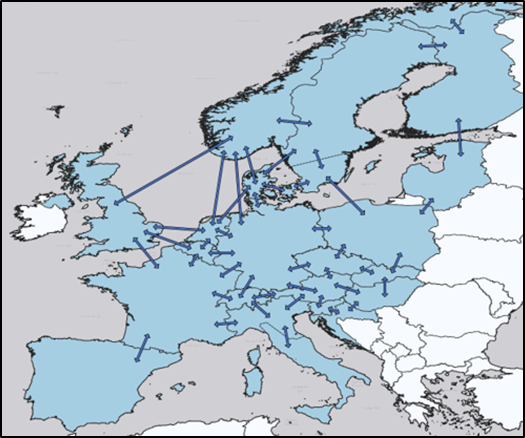}
  \caption{Geographical scope of the energy system model
  }
  \label{map_ESM}
\end{figure}

Our methodology can be applied to many liberalised electricity markets. We provide an empirical example for the German market. Due to the market integration of European electricity markets, we include most of the EU's 27 member states \footnote{Bulgaria, Cyprus, Greece, Iceland, Ireland, Malta and Romania are omitted.} as well as Norway, Switzerland, and the United Kingdom to include cross-border flows. Figure \ref{map_ESM} depicts the geographical scope of the \textit{em.power dispatch model} including all interconnections between the different market zones.

The model is specifically designed to generate hourly day-ahead electricity price forecasts. In contrast to many other energy system models, it uses a rolling window approach to clear the market optimally and uses exclusively input data known to market participants when submitting their bids. Further details about the proposed model can be found in \mycite{moebius2023enhancing}.

\subsection{Econometric Models}
\label{Met_econ}
This section will first describe the methodology of our state-of-the-art econometric model, the Ensemble Deep Neural Network model, and then introduce the econometric benchmark models.

\subsubsection{Ensemble Deep Neural Network Model}
\label{Met_EnsDNN}

The ensemble model category is one of the most commonly used forms of hybrid models aimed at overcoming the limitations of individual models and achieving highly accurate forecasting results. Theoretical and empirical evidence suggests that combining different models can significantly improve forecasting accuracy (\mycite{ben2018forecasting}, \mycite{niu2019combined}). Its primary aim is to leverage the advantages and strengths of individual models simultaneously. This conclusion is supported by \mycite{lago2021forecasting}, who compared the forecasting accuracy of state-of-the-art models, specifically Deep Neural Network (DNN) and the Lasso Estimated AutoRegressive (LEAR) models, with their ensemble versions. Their findings demonstrated that the ensemble models, namely Ensemble DNN (Ens-DNN) and Ensemble LEAR (Ens-LEAR), outperformed the individual models (i.e., DNN and LEAR, respectively).

Building on this finding, and recognising that the Ens-DNN model is the most accurate model (\mycite{lago2021forecasting}), we will consider it as the primary econometric component. This model will incorporate the MCP as an exogenous variable to test whether it benefits from the introduction of the output of the ESM model, further enhancing its forecasting accuracy.

Following \mycite{lago2021forecasting}, the Ens-DNN model is constructed by taking the arithmetic average of four distinct DNNs, each derived from running the hyperparameter and feature selection procedure four times. Hyperparameter optimisation, in theory, is a deterministic process over an infinite number of iterations, eventually identifying the global optimum. However, in practice, with a finite number of iterations and varying initial random seeds, the process becomes non-deterministic. This leads to different sets of hyperparameters and features each time it is run. These results, representing local minima, have nearly identical performance on the validation dataset, making it difficult to determine which is superior. This phenomenon occurs because deep neural networks (DNNs) are highly flexible, allowing multiple network architectures to achieve equally good results.
In this context, the combination of four distinct DNNs, each configured with unique hyperparameters, significantly enhances predictive accuracy compared to each DNN operating independently. The individual DNN model is explained in section \ref{Met_DNN}, where it is also introduced as an additional benchmark model.

\subsubsection{Econometric Benchmark Models}
\label{met_bench}

We use six additional econometric models to benchmark the results from the Ens-DNN: a DNN model, an Ensemble LASSO Estimated AutoRegressive (Ens-LEAR) model, the individual LEAR model, a Long-Short-Term Memory model (LSTM), a LASSO autoregressive with exogenous variables model (LARX), and a Random Forest model (RF).\\

\begin{figure}[h]
  \centering
  \includegraphics[width=0.9\textwidth]{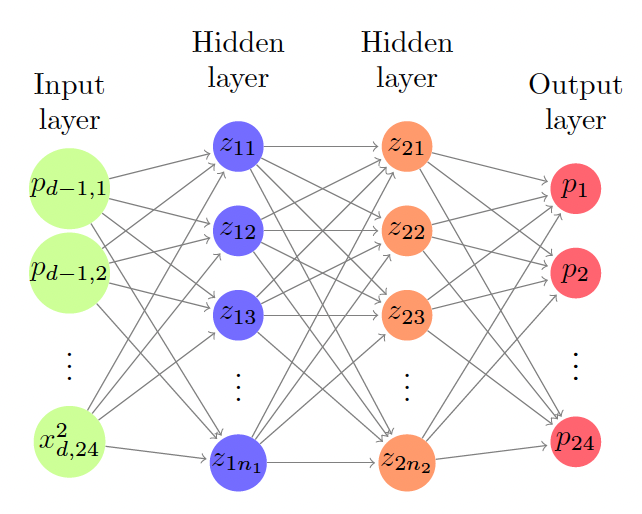}
  \caption{The DNN architecture adapted from \mycite{lago2018forecasting}}
  \label{dnn}
\end{figure}

\textbf{The Deep Neural Network Model} \\
\label{Met_DNN}
The DNN is a deep feedforward neural network with four layers and a multivariate framework (one model with 24 outputs). Without expert knowledge, its inputs and hyperparameters can be optimised for each case study. Figure \ref{dnn} shows the DNN model architecture.
To estimate the hyperparameters, the training dataset comprises data from the four years before the testing period. Throughout the training process, the training datasets are split into a training and a validation dataset, with the latter being used for two purposes: processing early stops to avoid overfitting and optimising hyperparameters. The validation dataset consists of 42 weeks.\\
Similar to the original DNN paper (\mycite{lago2018forecasting}), the tree-structured Parzen estimator (\mycite{bergstra2011algorithms}), a Bayesian optimisation approach based on sequential model-based optimisation, is used to jointly optimise the hyperparameters and input features. \\
To do this, the features are modelled as a set of hyperparameters, each of which is a binary variable that determines whether or not a particular feature is included in the model (\mycite{lago2018forecasting}). More specifically, the method uses 11 decision variables, or 11 hyperparameters, to determine which of the 241 possible input attributes are relevant:
\begin{itemize}
    \item {Four binary hyperparameters that determine whether or not the historical day-ahead prices $p_{d-1}$, $p_{d-2}$, $p_{d-3}$, and $p_{d-7}$ should be taken into account}
    \item{Two binary hyperparameters that specify whether or not to take into account the $\mathbf{x}^1_d$ and $\mathbf{x}^2_d$ day-ahead forecasts}
    \item{Four binary hyperparameters that determine whether or not the historical day-ahead forecasts $\mathbf{x}^1_{d-1}$, $\mathbf{x}^1_{d-7}$, $\mathbf{x}^2_{d-1}$, and $\mathbf{x}^2_{d-7}$ should be taken into account}
    \item{One binary hyperparameter that determines whether or not the variable $z_d$, which stands for the day of the week, should be used}
\end{itemize}
In other words, there are 10 binary hyperparameters that each determine whether or not to include 24 inputs. In addition, there is another binary hyperparameter that determines whether or not to include a dummy variable.\\ 
The technique additionally optimises eight more hyperparameters in addition to choosing the features: the number of neurons in each layer, the activation function, the learning rate, the dropout rate, the type of data preprocessing method, the use of batch normalisation, the initialisation weights of the DNN, and the coefficient for L1 regularisation that is applied to the kernel of each layer.

The hyperparameters and features are adjusted once using four years of data prior to the testing period, unlike the DNN weights, which are calibrated daily. It is crucial to remember that the algorithm runs for T iterations (T is set to 1500), during which time it infers a putative optimal subset of hyperparameters/features and assesses this subset on the validation dataset.\\

\textbf{The Ensemble LASSO Estimated AutoRegressive Model} \\
\label{EnsLEARmodel}

As for the Ens-DNN, the Ensemble LEAR model (Ens-LEAR) is constructed by taking the arithmetic average of individual forecasts derived from four different calibration window lengths. These lengths are specifically 1.5 years (78 weeks), 2 years, 3 years, and 4 years. This approach allows LEAR to incorporate a range of historical data perspectives, from short-term to long-term trends and cycles. With Ens-LEAR, forecasts are averaged over these varied time horizons, achieving a balance between recent data and more enduring historical patterning, and thus enhance prediction accuracy and robustness. Detailed information about the individual LEAR is provided in section \ref{LEARmodel}. It is also considered a benchmark.\\

\textbf{The LASSO Estimated AutoRegressive Model} \\
\label{LEARmodel}
LEAR is a parameter-rich Autoregressive ARX model estimated using LASSO as an implicit feature selection approach (\mycite{uniejewski2016automated}). The LEAR model is briefly outlined in this paragraph, with more comprehensive details provided in \mycite{lago2021forecasting}.\\
The model to predict electricity price $p_{d,h}$ on day $d$ and hour $h$ is given by:
\begin{equation}
    \begin{aligned}
p_{d, h}= & f\left(\mathbf{p}_{d-1}, \mathbf{p}_{d-2}, \mathbf{p}_{d-3}, \mathbf{p}_{d-7}, \mathbf{x}_d^i, \mathbf{x}_{d-1}^i, \mathbf{x}_{d-7}^i, \boldsymbol{\theta}_h\right)+\varepsilon_{d, h} \\
= & \sum_{i=1}^{24} \theta_{h, i} \cdot p_{d-1, i} \quad+\sum_{i=1}^{24} \theta_{h, 24+i} \cdot p_{d-2, i} \\
& +\sum_{i=1}^{24} \theta_{h, 48+i} \cdot p_{d-3, i}+\sum_{i=1}^{24} \theta_{h, 72+i} \cdot p_{d-7, i} \\
& +\sum_{i=1}^{24} \theta_{h, 96+i} \cdot x_{d, i}^1 \quad+\sum_{i=1}^{24} \theta_{h, 120+i} \cdot x_{d, i}^2 \\
& +\sum_{i=1}^{24} \theta_{h, 144+i} \cdot x_{d-1, i}^1+\sum_{i=1}^{24} \theta_{h, 168+i} \cdot x_{d-1, i}^2 \\
& +\sum_{i=1}^{24} \theta_{h, 192+i} \cdot x_{d-7, i}^1+\sum_{i=1}^{24} \theta_{h, 216+i} \cdot x_{d-7, i}^2 \\
& +\sum_{i=1}^7 \theta_{h, 240+i} \cdot z_{d, i} \quad+\varepsilon_{d, h}
\end{aligned}
\label{leareq}
\end{equation}
where $\boldsymbol{\theta}_h=\left[\theta_{h, 1}, \ldots, \theta_{h, 247}\right]^{\top}$ contains the LEAR model parameters for hour $h$ accounts of 247 parameters. As Equation \ref{leareq} is estimated using LASSO, many of these parameters become zero:

\begin{equation}
    \begin{aligned}
\hat{\boldsymbol{\theta}}_h & =\underset{\boldsymbol{\theta}_h}{\operatorname{argmin}}\left\{\operatorname{RSS}+\lambda\left\|\boldsymbol{\theta}_h\right\|_1\right\} \\
& =\underset{\boldsymbol{\theta}_h}{\operatorname{argmin}}\left\{\operatorname{RSS}+\lambda \sum_{i=1}^{247}\left|\theta_{h, i}\right|\right\}
\end{aligned}
\end{equation}
where $\text { RSS }=\sum_{d=8}^{N_d}\left(p_{d, h}-\hat{p}_{d, h}\right)^2$ is the sum of squared
residuals. $N_d$ is the number of days in the training sample, $\hat{p}_{d, h}$ is the price forecast, and $\lambda \geq 0$ is the regularisation hyperparameter of LASSO. \\
A hyperparameter that regulates the $L_1$ penalty is optimised upon recalibration every day. This can be accomplished using an ex-ante cross-validation procedure.\\

\textbf{Long-Short-Term Memory Model} \\
\label{lstmm}

\begin{figure}[]
  \centering
  \includegraphics[width=0.9\textwidth]{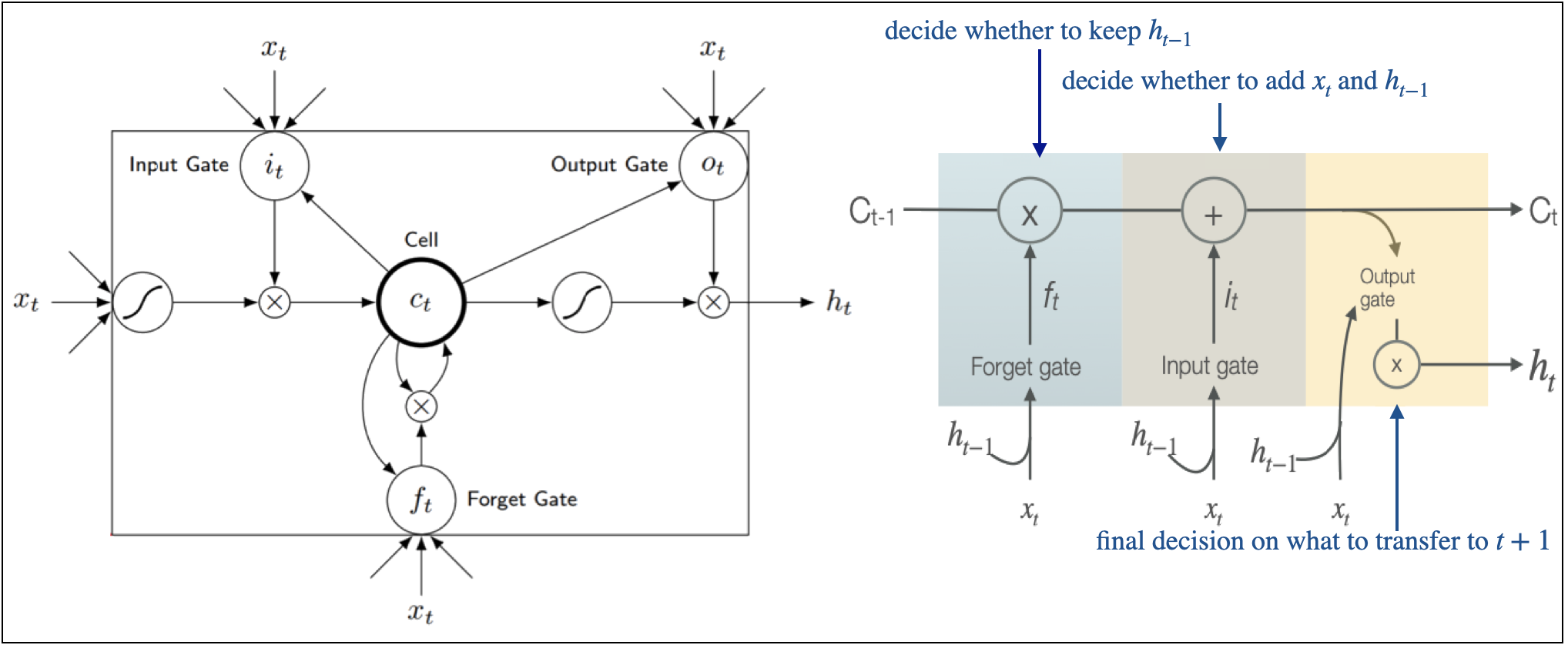}
  \caption{LSTM model architecture, adapted from \mycite{kawakami2008supervised}}.
  \label{lstm}
\end{figure}

One of the most crucial components of both current deep learning models and power market price forecasting is understanding long-term dependencies, which is the starting point for choosing and creating a forecasting model.
Like typical recurrent neural network models, the LSTM model has emerged as a crucial instrument for analysing complex feature analyses and sequence dependencies. The memory capacity of LSTM was greatly enhanced when compared to recurrent neural networks. More specifically, LSTM can handle long-term dependencies because it has built-in mechanisms that regulate how information is retained or forgotten over time. Due to their structure, these models are more effective at handling continuous sequences; as a result, they have a variety of applications in the field of EPF right now (\mycite{xiong2022hybrid}, and \mycite{li2021day}).\\
\mycite{hochreiter1997long} first proposed the LSTM architecture, which has since been improved by other researchers to yield superior performance (\mycite{shao2021feature}, and \mycite{shao2022pattern}).\\
The LSTM introduces the concept of "gates", specifically a forget gate, input gate, and output gate. As shown in Figure \ref{lstm}, in LSTMs, the forget gate controls how much information from the previous cell is retained, the input gate controls the input and update of the cell state, and the output gate computes the output. \
The forget gate $f_t$ is calculated at the $t$ time step based on the hidden state $h_t$ and the current input $x_t$ and then outputs data to multiply by the previous memory cell $C_{t-1}$. By using the forget gate $f_t$, data are output from $0$ to $1$, where $0$ represents completely forgotten data and $1$ represents completely reserved data. Thereafter, the input gate determines how much information can be transferred to the memory cell $C_t$. As part of the input gate process, the first step is to determine what information can be updated via sigmoid layers and the second step is to generate new information $C^*_t$ through the $tanh$ layers. Following that, each output from the forget gate and the input gate is used to calculate the current memory cell $C_t$. Finally, the output gate determines which information will be transmitted.\

The following are the equations used in the LSTM model:\
\begin{itemize}
\item{Input gate, $i_t$ at time $t$, and candidate cell state, $C^*_t$:}
\end{itemize}
\begin{equation}
i_t =\sigma\left(W_{x i} x_t+W_{h i} h_{t-1}+b_i\right) \\
\end{equation}
\begin{equation}
   C_t^*  =\tanh \left(W_{x c} x_t+W_{h c} h_{t-1}+b_c\right) 
\end{equation}
\begin{itemize}
\item{Activation of the memory cells' forget gate, $f_t$ at time $t$, and new state $C_t$:}
\end{itemize}
\begin{equation}
f_t  =\sigma\left(W_{x f} x_t+W_{h f} h_{t-1}+b_f\right) \\
\end{equation}
\begin{equation}
C_t  =f_t C_{t-1}+i_t C_t^*
\end{equation}

\begin{itemize}
\item{Activation of the cells' output gate, $o_t$, at time $t$, and their final outputs, $h_t$:}
\end{itemize}
\begin{equation}
o_t=\sigma\left(W_{x o} x_t+W_{h o} h_{t-1}+b_o\right) \\
\end{equation}
\begin{equation}
h_t=o_t \tanh \left(C_t\right)
\end{equation}
where:\\
$i_t$, $f_t$, and $o_t$ are the input gate, the forget gate, and the output gate, respectively,
$W_{xi}$, $W_{xc}$, $W_{xf}$, and $W_{xo}$ stand for the weight coefficients, 
$b_i$, $b_c$, $b_f$, and $b_o$ represent biases, 
$C^*_t$ represents the output in the input gate process, 
and $\sigma$ and $\tanh$ are activation functions.\\

\textbf{The LASSO Autoregressive Model With Exogenous Variables } \\
\label{larxm}

The autoregressive LASSO (Least Absolute Shrinkage and Selection Operator) technique suggested by \mycite{li2017sparse} is also adopted as a powerful statistical model. Statistical approaches have the advantage that their model complexity is much lower than that of ML methods, which means that they require fewer computational resources.\\
The use of autoregressive models in the prediction and inference of time series data is widespread. In light of this fact, we opted to adopt a linear autoregressive model with exogenous variables based on LASSO estimation. Throughout the manuscript, LARX is used to refer to this model. \\
The autoregressive model with exogenous variables is given by the following equation:
\begin{equation}
 y_{d,h}=\beta_0+\sum_{i=1}^I \beta_{h,i} x_{h,i}+\varepsilon_{d,h}
 \label{arx}
\end{equation}

where $y_{d,h}$ stands for the electricity prices of day $d$ and hour $h$, $x_i$ refers to the regressors stated in section \ref{regressors}, $\beta_i$ are their corresponding coefficients, and $\varepsilon_{d,h}$ are Gaussian variables, where $\varepsilon_{d,h} \sim N(0, 1)$. \\
The following notation can be used to write the compact matrix form of the model in Equation \ref{arx}:
\begin{equation}
\mathbf{y}=\mathbf{X} \boldsymbol{\beta}+\varepsilon
\end{equation}

In shrinkage (or regularisation), the coefficients of irrelevant explanatory variables are lowered towards zero so that the most relevant regressors are retained. In this way, the algorithm avoids overfitting and takes into account the most relevant regressors (\mycite{gareth2013introduction}).\\
A LASSO is a generalisation of linear regression that minimises the residual sum of squares (RSS) and a linear penalty function of the $\beta$s instead of just minimising the RSS:
\begin{equation}
 \hat{\boldsymbol{\beta}}_h^L=\min _\beta\left\{\mathrm{RSS}+\lambda\left\|\boldsymbol{\beta}_h\right\|_1\right\}=\min _\beta\left\{\mathrm{RSS}+\lambda \sum_{i=1}^n\left|\beta_{h, i}\right|\right\}
\end{equation}

where $\lambda \geq 0$ is the parameter that controls the level of sparsity in the model and is known as the tuning parameter (or regularisation parameter). It controls how many regressors are used to predict electricity prices (i.e. sparsity).
The bigger $\lambda$ is, the sparser the solution $\hat{\beta}_{LASSO}$ will be, which means that fewer data points from the input regressors will be used to predict the future.
 For $\lambda=0$, the model is reduced to a traditional AR model (Equation \ref{arx}). While in the case $\lambda \rightarrow \infty$ all coefficients $\beta_{h,i} \rightarrow 0$ tend towards zero, for intermediate values of $\lambda$ a balance must be struck between minimising RSS and shrinking coefficients towards zero. Practically, we estimate the value of $\lambda$ using $k$-fold cross validation (CV) (\mycite{hastie2009elements}). \\
In recent years, LASSO has become increasingly popular in EPF (\mycite{ludwig2015putting}, \mycite{ziel2015efficient}). As part of our paper, we analyse LASSO's ability to predict day-ahead electricity prices and provide a comparison with some other well-known prediction methods, i.e. LSTM model.\\
Compared to the LEAR model, LARX represents the foundational version from which \mycite{lago2021forecasting} has developed the state-of-the-art LEAR model. In our research, the use of both LARX and LEAR models offers insights into their distinct reactions under diverse input data. This comparative analysis serves as a valuable guide for market participants, allowing them to choose between the simpler LARX and the more complex LEAR models based on the characteristics of their data. \\

\textbf{The Random Forest RF Model} \\
\label{rfm}

The RF algorithm presented in this paper is based on bagging or bootstrap aggregation to tree learners. In this case, bagging involves repeatedly selecting $B$ bootstrap samples from the training set and fitting trees $T_b$ to these samples where $B$ is the number of trees and $T_b$ is a random forest tree.
For the model's parameter estimation, inputs $x=\left[x_1, \ldots, x_N\right]^T \in X$ and responses $y \in Y$ are introduced.
Following training, each tree's prediction is given. The predictors from all regression trees are averaged in the final prediction using the following formula:

\begin{equation}
    \hat{Y}_{r f}^B(x)=\frac{1}{B} \sum_{b=1}^B \hat{Y}_b(x)
\end{equation}
 where $\hat{Y}_b(x)$, $b=1,...B$ is the prediction of the $b_{th}$ RF tree.\\
 This model is often used for short-term EPF and is shown to provide accurate forecasts (\mycite{ludwig2015putting}, \mycite{portoles2018electricity}, and \mycite{mei2014random}).
 
\subsection{Model Combination}
The model combination is straightforward: we use the key output from the ESM, i.e. the market clearing price MCP, as an independent variable in the selected econometric model. To assess the value of this particular variable, we compare different models (ESM, Ens-DNN, and the six econometric benchmark models) as well as different data inputs in the econometric models: market fundamentals only, MCP only, and market fundamentals plus MCP. We thus have two dimensions of the analysis: first, the type of model, and second, the independent data entering the econometric models.

\subsection{Storage Optimisation}
\label{StorOpt}

After combining the ESM with the econometric models, we test the effectiveness of our price forecasts simulating the profit contribution of a storage application. Transporting energy over time, storage systems exploit price spreads and are therefore dependent on good price forecasts. To maximise their profits, they must first identify the time with the lowest electricity prices in order to charge their storage and then, at the time with the highest electricity prices, generate electricity and sell it back to the market.

To evaluate the effectiveness of a price forecast model, we generate a linear optimisation problem that maximises the profit contribution ($\Pi$) of a storage system, which results from purchases and sales of electricity at different hours ($h$) with different prices for a given number of days ($d$). 

The objective function of this problem is represented in Equation \ref{eqn_PC}. Using price forecasts ($\hat{p}_{d,h}$), optimal quantities and times for charging ($C_{d,h}$) and electricity generation ($G_{d,h}$) are determined. 

\begin{equation}
    \max \Pi = \sum_{d,h} \hat{p}_{d,h} \cdot G_{d,h} - \sum_{d,h} \hat{p}_{d,h} \cdot C_{d,h}
    \label{eqn_PC}
\end{equation}

The actual profit contribution ($\Pi^{act}$) of the storage system then results from the combination of the storage operations with the actual electricity prices ${p}_{d,h}$ (Equation \ref{eqn_PC_real}).

\begin{equation}
    \Pi^{act} = \sum_{d,h} {p}_{d,h} \cdot G_{d,h} - \sum_{d,h} {p}_{d,h} \cdot C_{d,h}
    \label{eqn_PC_real}
\end{equation}

Comparing this to the highest possible profit contribution, which is calculated using the actual electricity prices for determining optimal storage operations, we can derive the effectiveness of a price forecast.

Note that we focus on day-ahead price forecasts. Hence, we assume the storage only operates on the day-ahead market. Other streams of income such as intraday markets and control power markets are not taken into account. Additionally, we assume that the bidding strategy is based on the price forecasts for the following day only. To isolate the impact of daily price forecasts, the possibility of storing electricity for several days is neglected. Hence, the storage dispatch is determined daily before the day-ahead market closes at noon exclusively for the 24 hours of the following day. 

Both electricity generation ($G_{d,h}$) and consumption ($C_{d,h}$) are limited by the generation capacity ($cap$) of a storage system (Equation \ref{eqn_CapMax}). When analysing the relative effectiveness of a price forecast (i.e., the ratio of the realised profit contribution to the optimal profit contribution), the generation capacity can be set to one.

\begin{equation}
\begin{array}{l l}
    G_{d,h} + C_{d,h} \leq cap & \forall d\in D, h\in H
\end{array}
\label{eqn_CapMax}
\end{equation}

The possibility to generate or consume electricity depends on the current state of the storage level ($SL_{d,h}$) at the end of each hour. Equation \ref{eqn_SL} shows that withdrawing electricity from the market when charging increases the storage level, while generating electricity and selling it back to the market reduces it. Efficiency losses of the entire storage cycle are considered by the factor $\eta$, which is between zero and one. The energy storage must therefore charge more energy than it can generate later. Equation \ref{eqn_GenMax} states that the electricity generation must not be higher than the storage level at the end of the previous hour.

\begin{equation}
\begin{array}{l l}
    \ SL_{d,h} \leq  SL_{d,h-1} + C_{d,h} \cdot \eta - G_{d,h} & \forall d\in D, h\in H
\end{array}
\label{eqn_SL}
\end{equation}

\begin{equation}
\begin{array}{l l}
    \ G_{d,h} \leq  SL_{d,h-1} & \forall d\in D, h\in H
    \end{array}
\label{eqn_GenMax}
\end{equation}

A fully charged storage system can generate electricity for a certain number of full-load hours. This value is assumed to be exogenous in our analysis and is calculated as an energy-to-capacity ratio ($ecr$) restricting the storage level in Equation \ref{eqn_SLMax}.

\begin{equation}
\begin{array}{l l}
    SL_{d,h} \leq cap \cdot ecr & \forall d\in D, h\in H
    \end{array}
    \label{eqn_SLMax}
\end{equation}

As mentioned above, the storage operation is determined separately for each day. To avoid arbitrary effects, the storage status must be the same at the beginning and end of each day. We assume the storage to be empty at both times, which is controlled by Equations \ref{eqn_SL_1} and \ref{eqn_SL_24}.

\begin{equation}
\begin{array}{l l}
    SL_{d,h=1} =  C_{d,h} \cdot \eta & \forall d\in D
    \end{array}
    \label{eqn_SL_1}
\end{equation}

\begin{equation}
\begin{array}{l l}
    SL_{d,h=24} = 0 & \forall d\in D
    \end{array}
    \label{eqn_SL_24}
\end{equation}

Finally, the non-negativity constraints are presented in Equation \ref{eqn_NN}
\begin{equation}
\begin{array}{l l}
\ G_{d,h}, C_{d,h}, SL_{d,h} \geq 0 & \forall d\in D, h\in H
\end{array}
\label{eqn_NN}
\end{equation}

\section{Data and Model Preparation}
\label{data}

This section provides a structured overview of the dataset and input data used in developing the techno-economic energy system model and the econometric model for forecasting electricity prices in the German market.

\subsection{Input Data for the Energy System Model}
\label{fund_data}
We apply the European electricity market model \emph{em.power dispatch} with data from January 1st, 2015, until December 31st, 2020. In general, the use of ESMs requires extensive input data. In this section, we have listed the input data used to compute the MCP for the German electricity market.
On the \emph{demand side}, original load forecasts are taken from the ENTSOE transparency platform (\mycite{EntsoeTPe}). However, recent studies show that these data have structural errors that should be corrected (see, e.g. \mycite{Maciejowska2021}). Therefore, we run an error correction model for the day-ahead load forecasts according to (\mycite{moebius2023enhancing}). 

On the \emph{supply side}, a set of technologies is generating and storing electricity. Data on the technical-economic properties used in the model are listed in Table \ref{ESM_parameters}

\begin{table}[htbp]
\centering
\caption{Data used in the energy system model}
\begin{tabular}{ll}
\hline
			\multicolumn{1}{l}{\textbf{Parameter}} &
			\multicolumn{1}{l}{\textbf{Source}} \\
\hline
CO\textsubscript{2} prices	& \mycite{Sandbag2021}		\\
\hline
Control power procurement & \mycite{Regelleistung}	\\
\hline 
Curtailment costs for RES	& own assumption: 20 EUR/MWh		\\
\hline
Efficiency of generation capacities	&	\mycite{schroder2013current}, 	\\
 & \mycite{OPSDa}	\\
\hline
Efficiency losses at partial load	&	\mycite{schroder2013current} \\
\hline
Electricity demand \\(original day-ahead forecast)	& \mycite{EntsoeTPe}		\\
\hline
Energy-power factor (for storages) &	own assumption: 9	\\
\hline
Fuel prices	&	\mycite{Destatis2020},	\\
(Lignite, nuclear, coal, gas, oil)	&	\mycite{EEX2021}, \mycite{EntsoS} \\
\hline
Generation and storage capacity	&	\mycite{BNetzA2021}, \mycite{UBA2020}, \mycite{EBC2021},  \\
& \mycite{EntsoeTPa}, \\
&  \mycite{OPSDa}   \\
\hline
Generation by CHP units	&	\mycite{EC2021}	\\
\hline
Historic electricity generation	&	\mycite{EntsoeTPd}	\\
\hline
Load-shedding costs	&	own assumption: 3,000 EUR/MWh	\\
\hline
Minimum output levels  &  \mycite{schroder2013current}	\\
\hline
NTCs	&	\mycite{EntsoeTPf}, \\
&  \mycite{JAO2021}	\\
\hline
Variable O\&M costs  &  \mycite{schroder2013current}	\\
\hline
Power plant outages	&	\mycite{EntsoeTPb}	\\
\hline
RES generation	& \mycite{EntsoeTPc}		\\
\hline
Start-up costs	&	\mycite{schroder2013current}	\\
\hline
Seasonal availability of hydropower &   \mycite{EntsoeTPd}	\\
\hline
Temperature (daily mean)	&	\mycite{OPSDb}	\\
\hline
Water value 	&  \mycite{EntsoeTPd}, 	\\
& \mycite{EntsoeTPg}	\\
\hline
			\hline
\end{tabular}
\label{ESM_parameters}
\end{table}

The data provided for download on GitHub.\footnote{https://github.com/ProKoMoProject/Enhancing-Energy-System-Models-Using-Better-Load-Forecasts}

\subsection{Input Data for the Econometric Models}

We start this section with a description of the dependent variable, before turning to the independent variables entering the econometric model.

\subsubsection{Dependent Variable}

In this paper, we choose the hourly day-ahead wholesale electricity price for Germany as our target variable (dependent variable). This market is operated by EPEX SPOT. Our sample covers the period between January 1st, 2015, and December 31st 2020.

Day-ahead wholesale electricity trading is settled at 12 noon for each hour of the following day, meaning that all 24 hours of one day are traded at once. The forecasting process is thus divided into 24 separate daily prediction batches, with each batch corresponding to a specific hour of the day. This segmentation allows for the independent estimation of electricity prices for each hour, enabling a focused analysis of the unique factors influencing prices at different times of the day.

The electricity price data are sourced from the ENTSO-E Transparency Platform.
\subsubsection{Independent Variables}
\label{regressors}
A group of exogenous variables has been selected to forecast the wholesale electricity price. Our selection of independent variables is motivated by theoretical considerations (\mycite{kanamura2007structural}, \mycite{mount2006predicting}, \mycite{nogales2002forecasting}) and considers publicly available information. Six variables are considered: load forecasts, CO$_2$ emission prices, fuel prices (both gas and coal prices), as well as renewable generation (production forecasts for both wind and photovoltaic). 

Hourly load forecasts are included as independent variables in most articles on EPF (\mycite{lago2021forecasting}). As this paper focuses on day-ahead price forecasting, we use day-ahead load forecasts, improved with an error correction model (see \mycite{moebius2023enhancing}).\\
Prices for CO$_2$ emission certificates, hard coal, and natural gas are long-term variables that affect production costs, especially for the marginal plants determining prices with their variable costs. We use daily settlement prices for the 24 hours of the following day.\\
We include wind and solar generation forecasts as major renewable energy technologies in Germany.\\

Finally, the estimated market price from the techno-economic energy system model described in section \ref{fundm}, the MCP, is considered as an additional independent variable, which links the ESM and the econometric model.\\
The explanatory variables' sources are referred to in section \ref{fund_data}.\\

One point to notice is that the ESM has also taken into account the load forecast, the wind and solar generation, the gas and coal prices, and the CO$_2$ price. Through the demand balance constraint, these input data influence the values of other generation units. Econometric models, however, consider statistical patterns, such as volatility, long memory, and linear and non-linear trends, between electricity prices and the variables mentioned above. Although the same data were used in both models, they were not processed or treated the same way (\mycite{de2019electricity}).

\subsubsection{Training and Testing Periods}\

The training and testing samples were split up to evaluate the out-of-sample forecasting performance of our model. We use the first 4 years (the years between 2015 and 2018) as the training period, resulting in 34,680 hourly observations, and the following two years (2019 and 2020) are the testing period, resulting in 17,640 hourly observations. It is important to note that the length of both the testing and training periods was selected based on \mycite{lago2021forecasting}, who emphasised the importance of choosing two years to ensure effective research in EPF. 


\subsubsection{Preliminary Statistics} \

According to preliminary statistics, the spot price series exhibits typical characteristics of electricity prices, such as seasonality, high volatility, excess kurtosis, negative skewness, and spikes (see Table \ref{stat_Descp} and Figure \ref{pdf} in Appendix). \\
The test of stationarity (Augmented Dickey-Fuller ADF) rejected the unit root hypothesis at the 5\% significance level, indicating that the electricity price time series is stationary. 

\subsubsection{Feature Selection Algorithm}
\label{feature} \

To achieve the highest model performance, it is crucial to choose the optimum set of features from the original dataset. To avoid the curse of dimensionality, feature selection is usually an effective method. This process takes place when developing a predictive model by identifying a subset of attributes from the original dataset. It can reduce computation time, improve model predictions, and help us to get a better understanding of the dataset (\mycite{chandrashekar2014survey}, \mycite{li2021day}). Many feature selection methods have been proposed (\mycite{naz2019short}, \mycite{gholipour2018electricity}, \mycite{gao2018combination}). In this paper, we present a Recursive Feature Elimination (RFE) algorithm-based Random Forest (RF) model for selecting the most relevant features to develop an effective and reliable forecasting model (\mycite{chen2018decision}).\ 

\begin{figure}[h]
  \centering
  \includegraphics[width=0.6\textwidth]{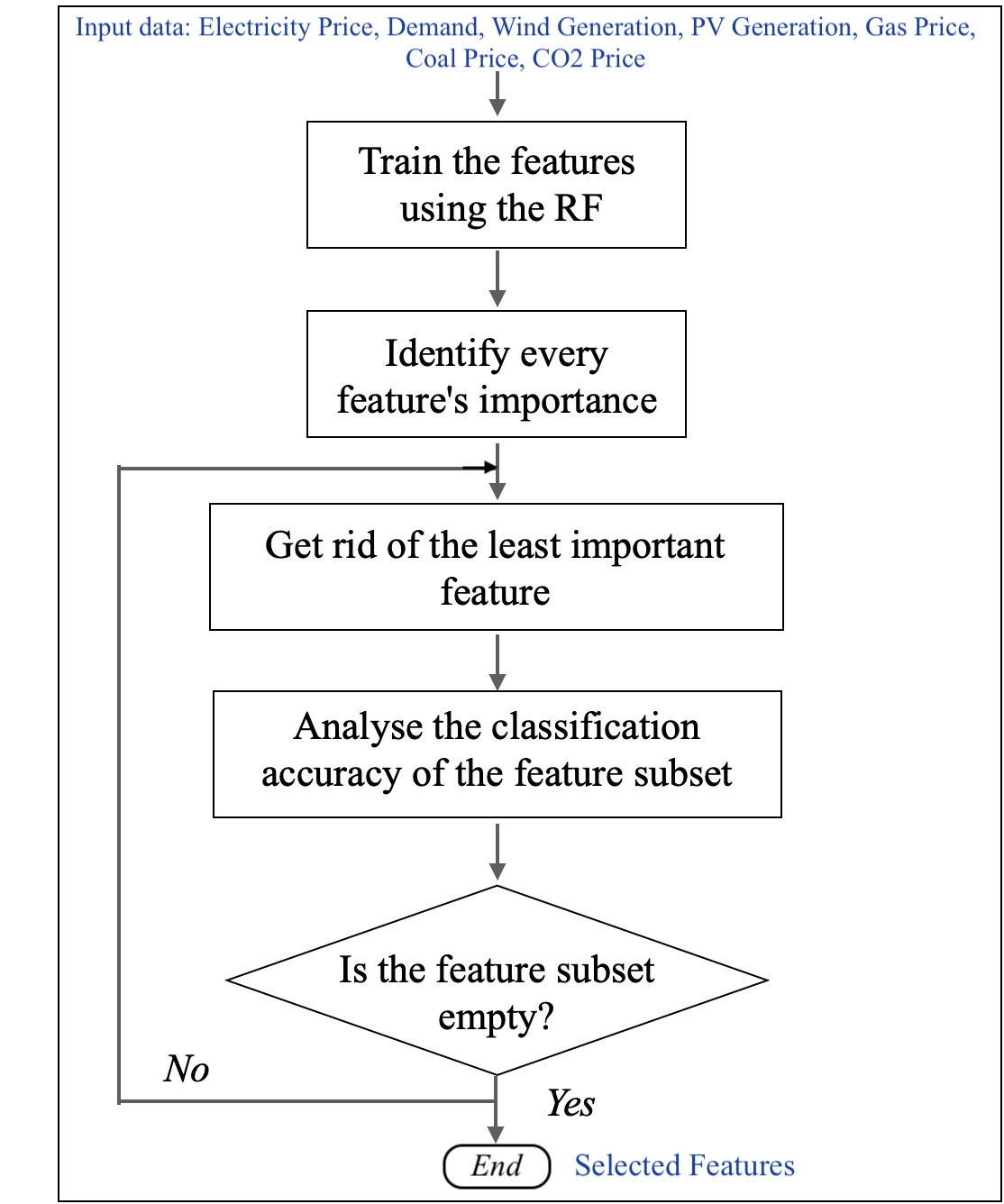}
  \caption{RF-RFE flowchart}
  \label{rfe}
\end{figure}

Figure \ref{rfe} depicts the RF-RFE approach's flowchart. First, we train our model using the RF algorithm with the training set. We then evaluate the significance of each feature based on its classification. Afterwards, the features are ranked by importance from the most important to the least important (see Figure \ref{feature}). Once the least relevant feature has been removed from the feature set, we retrain the RF model with the modified features and obtain the classification performance from the new feature set. As indicated by the name, RFE involves iterating until all crucial features are chosen. Several studies have demonstrated the effectiveness of this method (\mycite{liu2007computational}, \mycite{liu2005toward}).\\
Based on the feature selection results in Figure \ref{feature}, demand is the most important factor, followed by wind generation forecasts and gas prices. In the rest of the analysis, we keep all six exogenous variables since the algorithm does not exclude any of them.

\begin{figure}[h]
  \centering
  \includegraphics[width=0.9\textwidth]{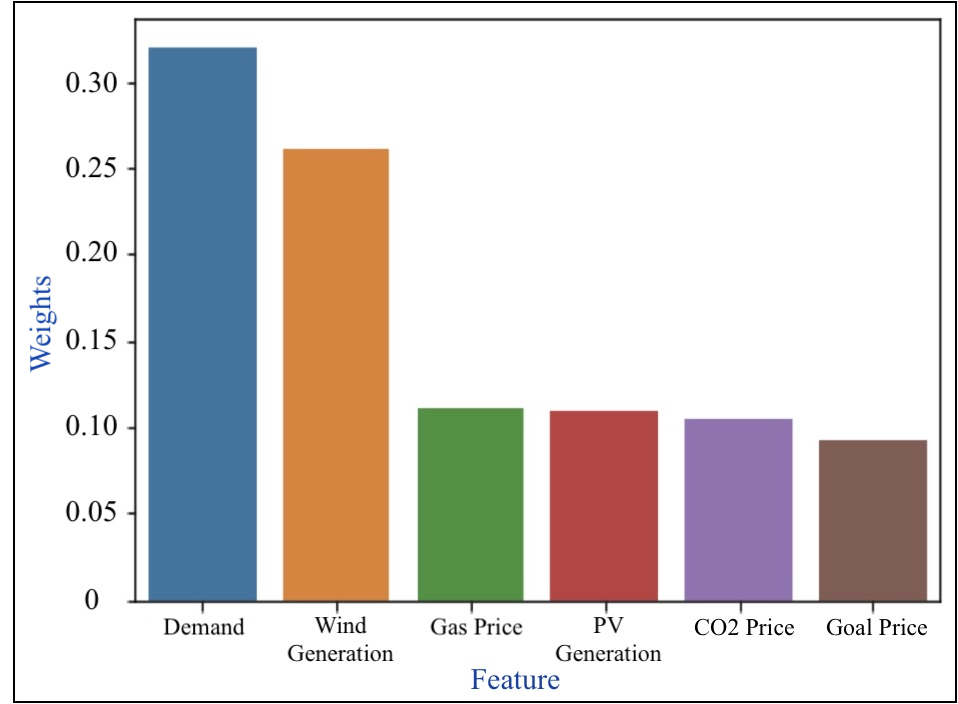}
  \caption{RFE-RF results}
  \label{feature}
\end{figure}

\subsection{Evaluation Metrics}
\label{evaluationsection}
The most widely used metrics to evaluate the accuracy of point forecasts in EPF are mean absolute error (MAE) and root mean square error (RMSE):
\begin{equation}
\mathrm{MAE} =\frac{1}{24 N_{\mathrm{d}}} \sum_{d=1}^{N_{\mathrm{d}}} \sum_{h=1}^{24}\left|p_{d, h}-\hat{p}_{d, h}\right|, \\
\end{equation}

\begin{equation}
\mathrm{RMSE}  =\sqrt{\frac{1}{24 N_{\mathrm{d}}} \sum_{d=1}^{N_{\mathrm{d}}} \sum_{h=1}^{24}\left(p_{d, h}-\hat{p}_{d, h}\right)^2,} \\
\end{equation}
where $p_{d,h}$ and $\hat{p}_{d,h}$ denote the real and forecasted prices on day $d$ and hour $h$, respectively, while the total number of days in the test period is denoted by $N_d$.\\
We also consider the symmetric mean absolute percentage error (sMAPE) as an alternative to the mean absolute percentage error (MAPE) \footnote{$
\mathrm{MAPE} =\frac{1}{24 N_{\mathrm{d}}} \sum_{d=1}^{N_{\mathrm{d}}} \sum_{h=1}^{24} \frac{\left|p_{d, h}-\hat{p}_{d, h}\right|}{\left|p_{d, h}\right|},
$}. The latter is not very informative, since it becomes very large at prices close to zero and is dominated by low prices. Taking into consideration that the day-ahead electricity price sample used in this study includes zero and close-to-zero prices, we adopt the sMAPE. The sMAPE is given by:

\begin{equation}
    \mathrm{sMAPE}=\frac{1}{24 N_{\mathrm{d}}} \sum_{d=1}^{N_{\mathrm{d}}} \sum_{h=1}^{24} 2 \frac{\left|p_{d, h}-\hat{p}_{d, h}\right|}{\left|p_{d, h}\right|+\left|\hat{p}_{d, h}\right|}
\end{equation}

The relative MAE (rMAE) is also considered, as a reliable evaluation metric that can overcome some limitations related to the mean absolute scaled error MASE, widely used in the literature. In particular, the MASE is dependent on the in-sample dataset, which implies that forecasting methods with different calibration windows will have to take into consideration different in-sample datasets. This results in the MASE of each model being based on a different scaling factor, so comparisons across models are not possible.
To overcome these limitations, the rMAE normalises MAE against a na\"ive forecast based on the out-of-sample dataset, ensuring consistent evaluation across models (\mycite{lago2021forecasting}). The rMAE is similar to MASE in the sense that rMAE normalises the MAE by the MAE of a na\"ive forecast. The rMAE is described as follows:
\begin{equation}
    \mathrm{rMAE}=\frac{\frac{1}{24 N_{\mathrm{d}}} \sum_{d=1}^{N_{\mathrm{d}}} \sum_{h=1}^{24}\left|p_{d, h}-\hat{p}_{d, h}\right|}{\frac{1}{24 N_{\mathrm{d}}} \sum_{d=1}^{N_{\mathrm{d}}} \sum_{h=1}^{24}\left|p_{d, h}-\hat{p}_{d, h}^{\text {naive }}\right|}
\end{equation}
where $\hat{p}_{d, h}^{\text {naive }}$ is the na\"ive forecast.

\section{Achieved Price Forecasting Accuracy}
\label{results}
This section evaluates the previously specified models for the forecasting of electricity prices. For the sake of clarity, we divide the section into two subsections.
The first subsection evaluates the performance of the techno-economic ESM, the Ens-DNN model, and their combinations, using the evaluation metrics defined in section \ref{evaluationsection}. The second subsection presents the comparative analysis to gauge the forecasting accuracy of the ESM--Ens-DNN model against alternative benchmark models incorporated within the combined framework instead of the Ens-DNN model. This serves two purposes: (i) it evaluates the effectiveness of the chosen Ens-DNN model as a candidate for the econometric model, and (ii) assesses how various statistical and machine learning models respond when integrated with the ESM, enabling us to evaluate the robustness and generalisability of improving econometric short-term price forecasts with results from the techno-economic ESM. 


\subsection{Performance of the ESM--Ens-DNN Model}
\label{dnnevsaluation}
The forecasting results in terms of evaluation metrics, the MAE, RMSE, sMAPE, and rMAE (described in section \ref{evaluationsection}), are presented in Table \ref{evaluationdnn}, with bold values representing the most favourable results.  

\begin{table}[h]
\caption{Forecasting results of the ESM--Ens-DNN model}
\resizebox{\textwidth}{!}{%
\begin{tabular}{llcccc}
\hline
\\
\textbf{Model} & \textbf{MAE} & \textbf{RMSE} & \textbf{sMAPE\%} & \textbf{rMAE} \\
\\
\hline
\\
ESM & 6.116 & 9.374 & 23.976 & 0.607 \\
\\
\text{Ens-DNN} & 4.272 & 7.222 & 19.180 & 0.457 \\
\\
\text{ESM--Ens-DNN +} & \textbf{3.496} & \textbf{5.907} & \textbf{16.660} & \textbf{0.374} \\
\\
\text{ESM--Ens-DNN} & 3.857 & 6.272 & 18.030 & 0.413 \\
\hline
\end{tabular}
}
\label{evaluationdnn}
\end{table}

All combinations of Ens-DNN and ESM are considered: first, the ESM is presented as a stand-alone forecasting model.\\
Second, the Ens-DNN model is used without incorporating the estimated price from the ESM. Six exogenous variables are considered in the Ens-DNN: a day-ahead forecast of demand, a wind generation forecast, and a PV generation forecast, as well as gas, coal, and CO$_2$ prices.\\
Third, the ESM is incorporated into the Ens-DNN through the addition of the MCP to the list of exogenous variables, denoted ESM--Ens-DNN+. Consequently, through a comparison between the stand-alone and combined models with exogenous variables, we can effectively highlight the pivotal role of including techno-economic information in enhancing the forecasting accuracy of the Ens-DNN model.\\
Fourth, we eliminate all exogenous variables and keep only the MCP, to test whether the ESM already contains all information about those variables. The model thus formed is denoted as ESM--Ens-DNN. This helps us to determine whether we can rely exclusively on the MCP derived from the ESM as a unique predictor of electricity prices. We can therefore test whether it is possible to reduce the number of features and thus reduce calculation time and costs without compromising prediction accuracy.\\
Looking at the table, several key take-aways can be derived. 
\newpage
\begin{result}[Three key take-aways on the Ens-DNN]
\
\begin{enumerate}
    \item The individual Ens-DNN outperforms the ESM across all evaluation metrics. This reconfirms the result that a (good) statistical model outperforms a (good) ESM in day-ahead forecasting.
    \item Model combination achieves a huge improvement in forecasting accuracy. The arithmetic average improvement from all evaluation metrics is approximately 18\% when comparing evaluation metrics for Ens-DNN and ESM--Ens-DNN+.
    \item Significant improvement can also be achieved with the ESM's MCP used as the only independent variable in the Ens-DNN. At the same time, the accuracy when the other six independent variables are added to the regression is even higher. The latter confirms the hypothesis stated in section \ref{regressors} that the ESM and the econometric model treat the same inputs differently (\mycite{de2019short}).\footnote{In the Ens-DNN, inputs are assigned a weight based on their relative importance. In the ESM, inputs affect the market clearing equilibrium by shaping the supply and demand side. Hence, these inputs influence the market price estimator based on economic theory in the ESM.} 
\end{enumerate}
\end{result}

 \begin{figure}[H]
  \centering
  \includegraphics[width=1\textwidth]{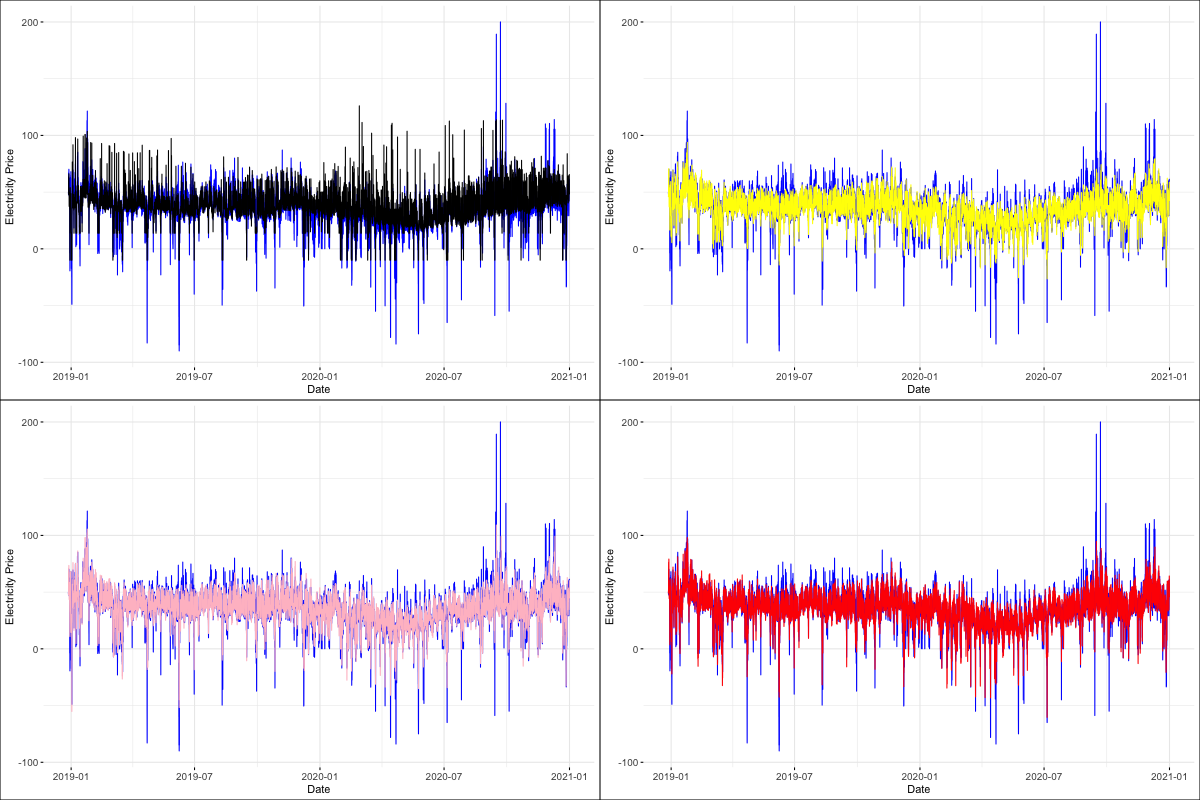}
  \caption{Comparison of \textcolor{blue}{real prices} and the ESM, \textcolor{yellow}{Ens-DNN}, \textcolor{pink}{ESM--Ens-DNN+}, and \textcolor{red}{ESM--Ens-DNN}.}
  \label{dnngraphs}
\end{figure}

Figure \ref{dnngraphs} plots the four different models we consider, i.e. the ESM and Ens-DNN plus the combined models ESM--Ens-DNN+ and ESM--Ens-DNN. The plots indicate that the forecasting of the Ens-DNN model is closer to the real data when compared to the ESM. Furthermore, the accuracy of the Ens-DNN model is further enhanced when combined with the ESM, as seen in the ESM--Ens-DNN+ and ESM--Ens-DNN models.\\

Overall, the results suggest that the combined model (in both versions: ESM--Ens-DNN+ and ESM--Ens-DNN) leverages the strengths of both the ESM and Ens-DNN models. On the one hand, the ESM simulates price formation from the intersection of demand and the physical and dynamic properties of the generators’ supply functions. On the other hand, the Ens-DNN model excels at capturing complex patterns and nonlinear relationships within data, making it particularly effective for short-term price forecasting tasks. By integrating these strengths, the combined model emerges as an excellent forecasting tool for short-term price predictions. 

Our key results can be confirmed when compared with the existing literature, where various models for electricity price forecasting have been developed. \mycite{ziel2018day}, for instance, forecast wholesale electricity prices with an overall MAE of 5.01 \euro{}/MWh for the years 2012 to 2016. \mycite{Maciejowska2021} achieve an RMSE of 8.43 \euro{}/MWh and a MAE of 5.92 \euro{}/MWh within the time period from 2016 to 2020 using an autoregressive model with exogenous variables. \mycite{marcjasz2023distributional} report a MAE of 3.542 and an RMSE of 6.146 for the German electricity market for the time horizon from June 26th, 2019, to December 31st, 2020, in the context of probabilistic forecasting.
Applying a techno-economic agent-based power market-simulation model, \mycite{qussous2022understanding} obtain an RMSE of 11.21 \euro{}/MWh and a MAE of 7.89 \euro{}/MWh for the years 2016 to 2019. 
Comparing our results to these studies, we can re-confirm that techno-economic models have relatively low forecasting accuracy. While our combination of ESM and Ens-DNN models outperforms the most accurate forecast by \mycite{marcjasz2023distributional}, it needs to be pointed out that the results of EPF models are difficult to compare in terms of input data and time horizons.

\subsection{Comparison with Benchmark Models and Robustness Check}
\label{evaluationbenchmark}

The proposed ESM--Ens-DNN model provides significant improvements in forecasting accuracy, both compared to the Ens-DNN alone as well as compared to the literature. However, to ensure the reliability and robustness of our findings, the methodology of using ESM results as input in regressions is tested on various econometric benchmark models, encompassing both statistical and machine learning approaches. The results are interpreted in the following section. The framework of the analysis (time horizon, independent variables, etc.) is identical to the previous subsection; the single variation is the chosen econometric model. 


The forecasting results of the evaluation metrics of the proposed ESM--Ens-DNN model against all the other model-data combinations are presented in Table \ref{evaluation1}. 

\begin{table}[H]
\centering
\caption{Benchmark models' forecasting results}
\setlength{\tabcolsep}{15pt} 
\begin{tabular}{lrrrrr}
\hline
\text{Model} & \text{MAE} & \text{RMSE} & \text{sMAPE} & \text{rMAE} \\
\midrule
\text{ESM} & 6.117 & 9.375 & 23.980 & 0.654 \\
\text{Ens-DNN} & 4.272 & 7.222 & \textbf{19.180} & 0.457 \\
\text{DNN} & 5.092 & 8.301 & 21.340 & 0.545 \\
\text{Ens-LEAR} & \textbf{4.090} & \textbf{6.852} & 19.860 & \textbf{0.438} \\
\text{LEAR} & 4.137 & 7.010 & 19.420 & 0.443 \\
\text{LSTM} & 6.073 & 8.655 & 24.840 & 0.650 \\
\text{LARX} & 8.576 & 10.977 & 29.880 & 0.917 \\
\text{RF} & 10.529 & 14.580 & 36.220 & 1.126 \\
\\
 \hline
 \\
\text{ESM--Ens-DNN +} & \textbf{3.496} & \textbf{5.907} & \textbf{16.660} & \textbf{0.374} \\
\text{ESM--DNN+} & 3.779 & 6.375 & 17.720 & 0.404 \\
\text{ESM--Ens-LEAR+} & 3.688 & 6.117 & 17.760 & 0.395 \\
\text{ESM--LEAR+} & 3.859 & 6.311 & 18.020 & 0.413 \\
\text{ESM--LSTM+} & 5.724 & 8.382 & 23.650 & 0.612 \\
\text{ESM--LARX+} & 10.743 & 12.683 & 34.300 & 1.149 \\
\text{ESM--RF+} & 11.808 & 16.628 & 38.330 & 1.263 \\
\\
 \hline
 \\
\text{ESM--Ens-DNN} & \textbf{3.857} & \textbf{6.272} & \textbf{18.030} & \textbf{0.413} \\
\text{ESM--DNN} & 3.986 & 6.300 & 18.400 & 0.426 \\
\text{ESM--Ens-LEAR} & 3.896 & 6.282 & 18.170 & 0.417 \\
\text{ESM--LEAR} & 4.160 & 6.627 & 19.000 & 0.445 \\
\text{ESM--LSTM} & 5.597 & 8.259 & 23.220 & 0.599 \\
\text{ESM--LARX} & 6.061 & 9.000 & 24.410 & 0.648 \\
\text{ESM--RF} & 8.129 & 12.677 & 29.210 & 0.869 \\
 \hline
\end{tabular}

\label{evaluation1}
\end{table}
\newpage
\begin{result}[Key take-aways for all models and robustness]
\
\begin{enumerate}
    \item Ens-LEAR outperforms the other models as stand-alone models. The difference is relatively small compared to Ens-DNN, LEAR, and DNN, considerable compared to LSTM and ESM, and large compared to LARX and especially RF.
    \item The best overall performance is achieved with the ESM--Ens-DNN+ model: it has minimal errors for all evaluation metrics. Hence, for all stakeholders exclusively interested in the most accurate day-ahead electricity price forecast, the ESM--Ens-DNN+ model is an excellent first candidate to implement.
    \item Use of the ESM's estimated market clearing price (MCP) as the only independent variable improves the forecasting quality of all econometric models. This confirms the robustness of our methodological approach, i.e. using the ESM's MCP as a single model input in an econometric model improves forecasting accuracy (lower third of the table).
\end{enumerate}
\end{result}

All in all, our result emphasises the strategic importance of the MCP as an exogenous variable in sophisticated econometric models. When the MCP is included alongside other independent variables, advanced models such as Ens-LEAR, Ens-DNN, LEAR, and DNN not only retain but also enhance their accuracy, demonstrating their robustness and compatibility with a complex variable set. In contrast, less sophisticated models like LARX and RF exhibit diminished accuracy when additional variables are included, indicating that their simpler structures might not effectively handle multiple regressors. Crucially, the Ens-LEAR, Ens-DNN, LEAR, and DNN models consistently outperform simpler models in all tested configurations.
This disparity underscores the significant advantage of employing more advanced models for integrating the MCP with other exogenous variables, aligning with our goal of optimising econometric model performance across various settings.

In detail, it is interesting to compare the Ens-DNN and the ESM-DNN, which indicates that the forecasting accuracy of the individual DNN, when introducing the MCP as a unique regressor, outperforms the ensemble DNN with the six exogenous variables. This highlights the importance of the MCP in enhancing the forecasting accuracy of the DNN model, making it outperform the four combined DNNs in its ensemble version with an average enhancement of 8\% across all evaluation metrics. Hence, investors can opt for the simpler ESM-DNN without sacrificing forecasting accuracy. For those seeking even higher precision, ESM-DNN+ offers superior results. If achieving the best forecasting outcome is the primary goal, investors should consider ESM--Ens-DNN+.

This range of models provides investors with the flexibility to select the most suitable one based on their cost-benefit trade-off, depending on their specific trading objectives.

Besides evaluation metrics, the error measures are statistically tested based on the Giacomini-White (GW) test. The results, presented in Appendix \ref{stat-tests}, confirm all our key take-aways from both sections \ref{dnnevsaluation} and \ref{evaluationbenchmark}. Hence, from a statistical perspective, the techno-economic ESM results indeed improve the econometric models' performance. 
 
\section{Achieved Improvements for Storage Bidding}
\label{storage}
In this section, we evaluate the practical benefit of our price forecasts. For this purpose, we assume a storage operator who buys electricity on the day-ahead market to fill a storage unit in order to withdraw and sell the electricity at a later time. The storage operator uses our price forecasts to select the time of storage at the lowest possible prices and the time of electricity withdrawal at the highest possible prices. Hence, the assumption is that storage dispatch decisions are made daily based on the available day-ahead price forecasts. In the second step, price realisations are revealed and storage revenues are calculated.\\
For the matter of generalisability, we apply three different storage types that differ in their energy-to-capacity ratio and storage cycle efficiency:
\begin{itemize}
\item Storage 1 has a rather high energy-to-capacity ratio of seven \footnote{An energy-to-capacity ratio of seven means that a full storage unit is able to generate electricity for seven straight hours at full load without being recharged} and an efficiency of 75~\%.
\item Storage 2 has a medium energy-to-capacity ratio of three and an efficiency of 80~\%.
\item Storage 3 has a low energy-to-capacity ratio of one and an efficiency of 90~\%.
\end{itemize}

\begin{figure}[H]
  \centering
  \includegraphics[width=1\textwidth]{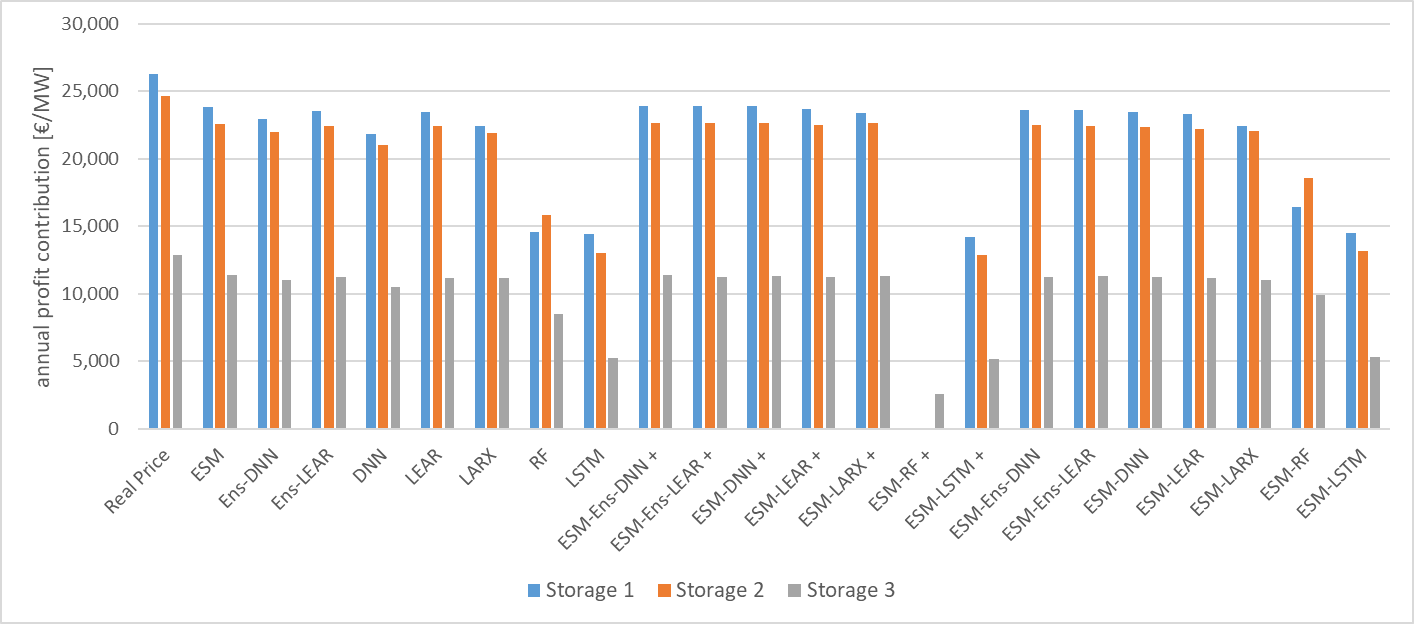}
  \caption{Annual profit contribution of storage plants in Euros per installed MW generation capacity}
  \label{apc}
\end{figure}

We present the optimisation model that determines the optimal storage dispatch in order to yield the highest possible profit contribution already in section \ref{StorOpt}. The model is fed with daily price forecasts made using the different forecasting models for 2019 and 2020. The day-ahead storage dispatch is decided based on these forecasts. Finally, the revenues of the storage unit are based on the revealed prices on the German wholesale electricity market. The code for this empirical application is provided on GitHub\footnote{https://github.com/BTU-EnerEcon/Bridging-ESM-and-Deep-Learning-Models}.\\
Note that we focus exclusively on day-ahead price forecasts in our study. Therefore, the storage units are optimised separately for each upcoming day and a cross-day storage deployment is not considered. Hence, the storage is given a fixed charge level at the beginning of each day. \\
Figure \ref{apc} shows the annual profit contribution per installed Megawatt of power generation capacity of the three storage units using all the forecast models presented in section \ref{method}.
A storage operator can achieve the highest possible return if a perfect forecast is available when planning the storage dispatch, i.e. the storage operator knows the actual price before the market clearing. The results of this hypothetical case are shown on the left side of the figure where the profit contribution is computed based on the actual day-ahead price (real price). This serves as a benchmark for the considered forecast models. \\
Per Megawatt installed capacity, storage with a high energy-to-capacity ratio generally achieves a higher profit contribution compared to storage with a lower energy-to-capacity ratio. This is because storage facilities with a high energy-to-capacity ratio can store electricity for several hours at a time and later generate electricity for several hours. For such units, a good forecast must identify not only the lowest and highest prices of a day but also the third-, fourth-, and fifth-lowest and highest prices.\\
Some of the applied models seem unable to meet these requirements, namely the RF and ESM--RF, as they do not capitalise on the value of a high energy-to-capacity ratio. The results also show that the ESM--RF+ model is generally not suitable for determining a storage dispatch. With prices ranging only between 29 and 34 EUR/MWh, the model does not identify sufficiently high price spreads to make storage worthwhile. \\
Table \ref{pc_factor} presents the profit contribution of storage as a factor to the perfect forecast, where the value of one would mean that the forecasting model achieves the results of a perfect forecast.\\
For all three storage types, we see a group of model types that achieve very good results that are relatively close to the optimal solution of a perfect forecast. These are the ESM, Ens-DNN, ESM--Ens-DNN+, ESM--Ens-DNN, Ens-LEAR, ESM--Ens-LEAR+, ESM--Ens-LEAR, LEAR, ESM--LEAR+, ESM--LEAR, ESM--DNN+, ESM--DNN, and ESM--LARX+.\\

\begin{result}[Key take-aways for storage bidding]
\
\begin{itemize}
    \item The ESM as a stand-alone model already achieves very good results. The ESM model is thus better - in comparison to econometric models - in identifying the optimal dispatch of storage than in forecasting day-ahead wholesale electricity prices. This may result from the ESM endogenously dispatching all capacities in the system (including storage) to determine the cost-minimal market clearing solution. Optimal storage dispatch is thus already part of ESM optimisation. Thus, the model is stronger in identifying "scarcity" hours when storage should produce and "excess" hours when storage should charge than in deriving the absolute level of prices.
    \item Revenues increase for all econometric models when the ESM's MCP is added as a regressor. Eventually, the ESM--DNN+ model, which combines a deep neuronal network with an energy system model, yields the highest profit contribution for Storage 1 and Storage 2, closely followed by the ESM and ESM--LEAR models. For Storage 3, the stand-alone ESM is the most profitable forecasting model, closely followed by the Ens-DNN+, the ESM--LARX+ and ESM--DNN+, and the ESM--Ens-LEAR models.
    \item These two results in combination, i.e. first that the ESM alone provides near-optimal storage revenues, and second that the econometric models benefit from including the MCP as an independent variable, re-confirm the high practical value of the techno-economic ESM in short-term price forecasting. On the econometric side, the ESM--Ens-DNN+ model combination is found to be an excellent model to generate accurate price forecasts as well as optimise storage operation (and maximise storage revenues) for all considered storage types.   
\end{itemize}
\end{result}

\begin{table}[H]
\centering
\caption{Profit contribution of storage plants as a factor to a perfect forecast}
\begin{tabular}{llll}
\hline
				    &	Storage 1	&	Storage 2 &	Storage 3		\\
\hline
ESM	&	0.907	&	0.916	&	\textbf{0.888} \\
Ens--DNN &	0.875	&	0.891	&	0.858 \\
Ens--LEAR &	0.896	&	0.909	&	0.874 \\
DNN 	&	0.831	&	0.852	&	0.817 \\
LEAR 	&	0.893	&	0.909	&	0.870 \\
LARX	&	0.853	&	0.888	&	0.868 \\
RF 	&	0.555	&	0.643	&	0.665 \\
LSTM 	&	0.549	&	0.528	&	0.411 \\

\\
 \\
ESM--Ens-DNN+ &	\textbf{0.911}	&	0.918	&	0.887 \\
ESM--Ens-LEAR+ &	\textbf{0.911}	&	0.917	&	0.878 \\
ESM--DNN+	&	0.910	&	\textbf{0.919}	&	0.880 \\
ESM--LEAR+ 	&	0.902	&	0.913	&	0.876 \\
ESM--LARX+	&	0.892	&	0.917	&	0.881 \\
ESM--RF+	&	0.000	&	0.000	&	0.201 \\
ESM--LSTM+	&	0.542	&	0.521	&	0.403 \\

\\
 \\
ESM--Ens-DNN &	0.899	&	0.911	&	0.878 \\
ESM--Ens-LEAR &	0.899	&	0.909	&	0.880 \\
ESM--DNN 	&	0.894	&	0.906	&	0.873 \\
ESM--LEAR 	&	0.889	&	0.900	&	0.868 \\
ESM--LARX 	&	0.842	&	0.866	&	0.812 \\
ESM--RF 	&	0.626	&	0.751	&	0.770 \\
ESM--LSTM 	&	0.553	&	0.534	&	0.416 \\

\hline

\end{tabular}
\label{pc_factor}
\end{table}

\newpage
\section{Conclusion}
\label{conclusion}

Our study combines two fundamentally different state-of-the-art price forecasting models for electricity prices: a techno-economic energy systems model and an econometric Ens-DNN model. The purpose is, first, to predict day-ahead electricity prices in Germany with high accuracy and, second, to assess the value of ESMs in short-term price forecasting. Our approach is robustified with a strategic selection of additional econometric models, encompassing a spectrum of statistical and machine learning models with distinct characteristics. We included linear and nonlinear models, ensemble learning (decision trees) models, and statistical state-of-the-art models. This deliberate diversity allows us to thoroughly evaluate how these models respond to the combination with an ESM, thereby assessing the robustness and generalisability of our proposed methodology. Our research thus compares forecast accuracy caused by changes in two dimensions: the type of econometric model adopted, and the set of the independent variables used in the econometric models (fundamental regressors without the ESM's market clearing price estimator, fundamental regressors plus the ESM's market clearing price estimator, and the ESM's market clearing price estimator as the only regressor).

Our analysis provides numerous insights. First, we re-confirm that techno-economic ESMs alone are better suited for medium- to long-term forecasts and cannot play to their strengths when directly forecasting high-frequency day-ahead prices. Standing alone, most econometric models are more accurate when parameterised with six fundamental independent variables. Second, we find that the picture pivots by 180° when ESM results are used as an alternative input in econometric models: all seven econometric models considered become more accurate when the fundamental regressors are replaced by the ESM's market clearing price. This result proves that ESM results can improve electricity price forecasting accuracy considerably when used in combination with econometric models.  

Third, the highest overall EPF accuracy is achieved in the ESM--Ens-DNN+ model, i.e. when the Ens--DNN model is run with the six fundamental independent variables plus the ESM's MCP. In fact, the five most accurate econometric models achieve the highest forecasting in this combination of regressors.\footnote{Exceptions are the LARX and RF models, which show better performance when the MCP is used as the sole regressor. This result, as well as the result that no further improvement is achieved in these two models when the MCP is added to the other regressors, can be due to their limited feature selection capabilities, which suggests that these models benefit from a simplified input structure where a strong single predictor dominates, avoiding potential noise from weaker variables.} In the case of the ESM--Ens-DNN+, the accuracy outperforms the literature cited in this paper, but we have pointed out that results of EPF models are difficult to compare as they vary in terms of input data and time horizons.  

From a practical standpoint, we have demonstrated the economic impact of our forecasting models through the optimisation of an electricity storage resource. Based on our forecasts, the operator optimised storage and withdrawal timings to maximise the profit contributions. Overall, the most accurate price forecasts also deliver the highest profit contributions, in particular the ESM--Ens-DNN+. A notable exception is the individual ESM, which is stronger in deciding the optimal dispatch of the storage than it is in forecasting exact prices. Our analysis provides investors with a diverse set of models to choose from. In particular, we allow them to balance forecast accuracy with profit objectives and model complexity, effectively balancing the cost-benefit trade-offs of each forecasting model. With this flexibility, investors can align their financial goals with market conditions and make informed decisions.\

Although we employ state-of-the-art models and other models with unique strengths and characteristics, there remain a variety of powerful econometric models that are worth testing further by applying our methodology. This is especially pertinent given the rapidly growing body of literature on electricity price forecasting including probabilistic methods (\mycite{marcjasz2023distributional}) and hybrid models, which typically combine data decomposition algorithms, statistical and machine learning models, and optimisation methods (\mycite{ehsani2024price}). One potential future research direction could involve testing our methodology across various econometric models and multiple datasets.

\section*{Acknowledgements}
Souhir Ben Amor gratefully acknowledges the support of the German Federal Ministry for Economic Affairs and Climate Action (BMWK), which funded this research through the FOCCSI 2 project (Forecast Optimization by Correction and Combination Methods for System Integration; Award No. 03EI1061) as part of the 7th Energy Research Program. Thomas Möbius thanks the German Federal Ministry of Economic Affairs and Climate Action through the research project ProKoMo, Award No. 03ET4067A within the Systems Analysis Research Network of the 6th energy research program. Felix Müsgens gratefully acknowledges financial support from the Federal Ministry of Education and Research, Award No. 19FS2032C, as well as the German Federal Government, the Federal Ministry of Education and Research, and the State of Brandenburg within the framework of the joint project EIZ: Energy Innovation Center (project numbers 85056897 and 03SF0693A) with funds from the Structural Development Act (Strukturstärkungsgesetz) for coal-mining regions. 


\printbibliography

\section{Appendix}

\subsection{Descriptive Statistics}
\begin{table}[H]
	\begin{minipage}{0.5\linewidth}
		\caption{Descriptive statistics for the German electricity price}
		\label{table:student}

		\begin{tabular}{|ll|}
\hline
\multicolumn{2}{|l|}{Electricity price}                                                                  \\ \hline
\multicolumn{1}{|l|}{Observations}      & 52.608                                                         \\ \hline
\multicolumn{1}{|l|}{Mean}              & 34.563                                                         \\ \hline
\multicolumn{1}{|l|}{std}               & 16.608                                                         \\ \hline
\multicolumn{1}{|l|}{min}               & -130.090                                                       \\ \hline
\multicolumn{1}{|l|}{25\%}              & 25.920                                                         \\ \hline
\multicolumn{1}{|l|}{50\%}              & 34.020                                                         \\ \hline
\multicolumn{1}{|l|}{75\%}              & 43.590                                                         \\ \hline
\multicolumn{1}{|l|}{Max}               & 200.040                                                        \\ \hline
\multicolumn{1}{|l|}{kurtosis}          & 7.808                                                          \\ \hline
\multicolumn{1}{|l|}{skewness}          & -0.216                                                         \\ \hline
\multicolumn{1}{|l|}{Jarque\_bera test} & \begin{tabular}[c]{@{}l@{}}81047.046\\ (0.000)***\end{tabular} \\ \hline
\multicolumn{1}{|l|}{ADF test}          & \begin{tabular}[c]{@{}l@{}}-15.288\\ (0.000)*** \end{tabular}   \\ \hline
\end{tabular}
\label{stat_Descp}
	\end{minipage}\hfill
	\begin{minipage}{0.6\linewidth}
		\centering
		\includegraphics[width=90mm]{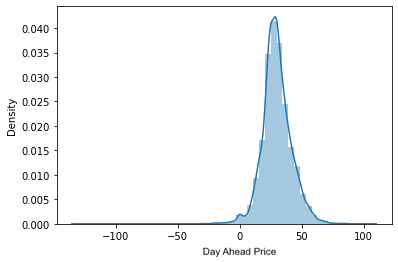}
		\captionof{figure}{Density function}
		\label{pdf}
	\end{minipage}
 \footnotesize{Levels of the significance of Jarque-Bera and ADF\\
 tests are indicated between squared brackets.\\
*** Denotes significance at the 1\% level.}\\
\end{table}

\subsection{Robustness Based on Statistical Testing}
\label{stat-tests}
Besides evaluation metrics, it is important to determine whether any difference in accuracy is statistically significant. To determine whether the forecast accuracy difference is real and not just due to random differences between forecasts, it is crucial to perform statistical tests. It has been argued that the Giacomini-White (GW) test is preferable because it can be viewed as a generalisation of the Diebold-Mariano (DM) test (the reader may refer to \mycite{lago2021forecasting} for more details).\\
We perform the multivariate GW test jointly for all hours by using the multivariate loss differential series or the daily loss differential series:
\begin{equation}
  \Delta_d^{\mathrm{A}, \mathrm{B}}=\left\|\varepsilon_d^{\mathrm{A}}\right\|_p-\left\|\varepsilon_d^{\mathrm{B}}\right\|_p,
  \label{eqmultivariateDM}
\end{equation}
where $\varepsilon_{d, h}^{\mathrm{A}}=p_{d, h}-\hat{p}_{d, h}$ is the prediction error of
model $A$ for day $d$ and hour $h$.\\
According to the results of the multivariate GW test using the $L_1$ norm in Equation \ref{eqmultivariateDM}, we have the following loss differential series:
\begin{equation}
    \Delta_d^{\mathrm{A}, \mathrm{B}}=\sum_{h=1}^{24}\left|\varepsilon_{d, h}^{\mathrm{A}}\right|-\sum_{h=1}^{24}\left|\varepsilon_{d, h}^{\mathrm{B}}\right|
    \label{GW}
\end{equation}

\begin{figure}[h]
  \centering
  \includegraphics[width=1\textwidth]{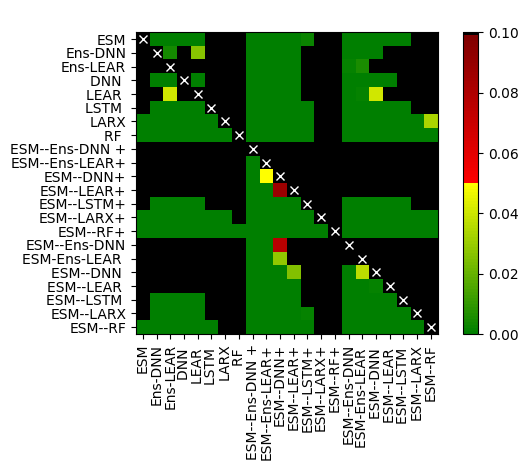}
  \caption{Results of the GW test}
  \label{DW}
\end{figure}

In Figure \ref{DW} we display the results for all models.\\

 To illustrate the range of obtained $p-values$, we use heat maps arranged as chessboards. Given that this test is run at a 5\% significance level, results are interpreted based on $p-values$ as follows:
\begin{itemize}
\item {$p-values$ $<$ 0.05 (green) indicate significant outperformance of a model on the X-axis (better) compared to the model on the Y-axis (worse).}
\item{$p-values$ $>$ 0.05 indicate significant underperformance of the model on the X-axis (worse) compared to the model on the Y-axis (better).}
\end{itemize}

The following results were withdrawn:
\begin{itemize}

\item{The ESM--Ens-DNN+ column is green. This indicates that this model's predictions are statistically significantly better than those of all the other models.}
\item{The forecasts of all econometric models combined with the ESM are statistically significantly better than their stand-alone versions. This shows the importance of the ESM's MCP in increasing the forecasting accuracy of the econometric models.}
\end{itemize}

From a statistical standpoint, the MCP contribution of the ESM enhances the performance of individual econometric models. Moreover, by examining the results of models both with and without independent variables, in addition to the MCP (e.g., ESM-DNN+ vs ESM-DNN), we can observe how different models respond to the inclusion of extra predictors. This interaction allows us to differentiate between models that perform better with a larger number of features and those that excel with a few, but more potent, features.

\end{document}